\documentclass[]{spie}  

\usepackage{amsmath,amsfonts,amssymb}
\usepackage{graphicx}
\usepackage{siunitx}
\usepackage[colorlinks=true, allcolors=blue]{hyperref}

\title{Laboratory Measurements of Instrumental Signatures of the LSST Camera Focal Plane}

\author[a]{Adam Snyder}
\author[e]{Aur\'{e}lien Barrau}
\author[bc]{Andrew Bradshaw}
\author[c]{Boyd Bowdish}
\author[bc]{James Chiang}
\author[d]{Celine Combet}
\author[bc]{Seth Digel}
\author[bc]{Richard Dubois}
\author[d]{Ludovic Eraud}
\author[e]{Claire Juramy}
\author[a]{Craig Lage}
\author[c]{Travis Lange}
\author[d]{Myriam Migliore}
\author[f]{Andrei Nomerotski}
\author[f]{Paul O'Connor}
\author[f]{HyeYun Park}
\author[bc]{Andrew Rasmussen}
\author[bc]{Kevin Reil}
\author[bc]{Aaron Roodman}
\author[g]{Adrian Shestakov}
\author[bc]{Yousuke Utsumi}
\author[g]{Duncan Wood}

\affil[a]{Department of Physics and Astronomy, University of California/Davis, Davis, California, United States}
\date{October 2020}
\affil[b]{Kavli Institute for Particle Astrophysics and Cosmology, Stanford University, Stanford, California, United States}
\affil[c]{SLAC National Accelerator Laboratory, Menlo Park, California, United States}
\affil[d]{Laboratoire de Physique Subatomique et de Cosmologie, Universit\'{e} Grenoble-Alpes, CNRS/IN2P3, Grenoble, France}
\affil[e]{LPNHE, (CNRS/IN2P3, Sorbonne Universit\'{e}, Paris Diderot), Laboratoire de Physique Nucl\'{e}aire et de Hautes \'{E}nergies, Paris, France}
\affil[f]{Brookhaven National Laboratory, Upton, New York, United States}
\affil[g]{Department of Physics, University of California/Santa Cruz, Santa Cruz, California, United States}

\authorinfo{Send correspondence to Adam Snyder: E-mail: aksnyder@ucdavis.edu}

\begin{document}

\maketitle

\begin{abstract}
Electro-optical testing and characterization of the Vera C. Rubin Observatory Legacy Survey of Space and Time (LSST) Camera focal plane, consisting of 205 charge-coupled devices (CCDs) arranged into 21 stand-alone Raft Tower Modules (RTMs) and 4 Corner Raft Tower Modules (CRTMs), is currently being performed at the SLAC National Accelerator Laboratory.  Testing of the camera sensors is performed using a set of custom-built optical projectors, designed to illuminate the full focal plane or specific regions of the focal plane with a series of light illumination patterns: the crosstalk projector, the flat illuminator projector, and the spot grid projector.  In addition to measurements of crosstalk, linearity and full well, the ability to project realistically-sized sources, using the spot grid projector, makes possible unique measurements of instrumental signatures such as deferred charge distortions, astrometric shifts due to sensor effects, and the brighter-fatter effect, prior to camera first light.  Here we present the optical projector designs and usage, the electro-optical measurements and how these results have been used in testing and improving the LSST Camera instrumental signature removal algorithms.
\end{abstract}

\section{INTRODUCTION}\label{sec:introduction}

The Vera C. Rubin Observatory, currently under construction on Cerro Pach\'{o}n in northern Chile, is an 8-meter-class telescope that will conduct the 10-year Legacy Survey of Space and Time (LSST) using the 3.2 gigapixel LSST Camera. The four main science goals of this future wide, fast, and deep survey are to: study the nature of dark matter and dark energy, create a detailed catalog of the solar system, explore the transient optical sky, and study the structure and formation of the Milky Way.\cite{Ivezic2008} The combination of a large field of view (\ang{3.5}) and fast cadence (a new field every 30 seconds) will allow the LSST Camera to image nearly a quarter of the sky in a single filter every three nights, resulting in an unprecedented amount of data that will be collected during the survey.

The LSST Camera focal plane has a diameter of approximately 64 cm and is made up of a mosaic of 205 charge-coupled devices (CCDs) arranged into 21 science raft tower modules (RTMs) and 4 specialized Corner Raft Tower Modules (CRTMs) for wavefront sensing and guiding. An individual RTM contains all the required thermal, mechanical, and electronic connections needed to operate the CCDs and can itself be considered a self-contained camera. In addition to the science RTMs, there are four specialized corner raft tower modules (CRTMs) used for telescope guiding and to provide wavefront sensing capabilities for the telescope's active optics system. The custom-designed 16 megapixel LSST Camera CCDs, supplied by Imaging Technology Laboratories (ITL) and Teledyne e2v, are fully depleted high-resistivity bulk silicon sensors that are segmented into 16 separate sections, each with their own output amplifier electronics. In order for the Rubin Observatory project to achieve its stated science goals to the desired precision and accuracy, each CCD must meet a set of stringent electro-optical requirements (Table \ref{tab:ccd_requirements}), determined by a number of past studies \cite{Radeka2009, Doherty2014, Kotov2016}. Additional sensor effects that must be characterized and, if necessary, corrected via hardware modifications or software algorithms in data reduction include source astrometric and shape distortions caused by pixel-area variations, electronic crosstalk, the brighter-fatter effect, and deferred charge during CCD readout. 

\begin{table}[ht]
\caption{Table of electrical and optical requirements for the LSST Camera CCDs \cite{Radeka2009, Doherty2014, Kotov2016}.} 
\label{tab:ccd_requirements}
\begin{center}       
\begin{tabular}{|c|c|} 
\hline
\rule[-1ex]{0pt}{3.5ex} Description & Specification \\
\hline\hline
\rule[-1ex]{0pt}{3.5ex} Read noise & $<8\,\, e^-$ \\
\hline
\rule[-1ex]{0pt}{3.5ex} Blooming full-well & $<175$ k$e^-$ \\
\hline
\rule[-1ex]{0pt}{3.5ex} Non-linearity & $< \pm 2$\% \\
\hline
\rule[-1ex]{0pt}{3.5ex} Serial CTI & $<5.0 \times 10^{-6}$ \\
\hline
\rule[-1ex]{0pt}{3.5ex} Parallel CTI & $<3.0 \times 10^{-6}$ \\
\hline
\rule[-1ex]{0pt}{3.5ex} Crosstalk & $<0.0019$ \\
\hline
\rule[-1ex]{0pt}{3.5ex} Dark Current 95th Percentile & $<0.2$ electrons per sec \\
\hline
\rule[-1ex]{0pt}{3.5ex} u-band Quantum Efficiency & $>41$\% \\
\hline
\rule[-1ex]{0pt}{3.5ex} g-band Quantum Efficiency & $>78$\% \\
\hline
\rule[-1ex]{0pt}{3.5ex} r-band Quantum Efficiency & $>83$\% \\
\hline
\rule[-1ex]{0pt}{3.5ex} i-band Quantum Efficiency & $>82$\% \\
\hline
\rule[-1ex]{0pt}{3.5ex} z-band Quantum Efficiency & $>75$\% \\
\hline
\rule[-1ex]{0pt}{3.5ex} Y-band Quantum Efficiency & $>21$\% \\
\hline
\rule[-1ex]{0pt}{3.5ex} Pixel Response Non-uniformity & $<5.0$\% \\
\hline
\rule[-1ex]{0pt}{3.5ex} Point Spread Function & $\sigma<5.0$ $\mu$m \\
\hline
\end{tabular}
\end{center}
\end{table}

There are three levels of testing for each of the focal plane CCDs: sensor-level acceptance testing, raft-level acceptance testing, and focal plane testing following integration of the RTMs into the cryostat.  Sensor acceptance testing was performed in three steps prior to the assembly of the RTMs: acceptance testing by the CCD vendors, reprocessing of the vendor data at SLAC, and acceptance testing at the Brookhaven National Laboratory (BNL) using a single sensor cryostat and commercial electronics controller\cite{Kotov2016}. Raft acceptance testing was first performed at BNL after assembly of the RTMs, before each RTM was shipped to the SLAC National Accelerator Laboratory where the integration and testing of the LSST Camera is being performed. After receipt, the RTMs were re-verified before being safely stored to await installation into the cryostat and focal plane testing\cite{Lopez2018}.

Section \ref{sec:focal_plane_testing} begins with a brief summary of the LSST Camera focal plane testing equipment and procedures used at SLAC. Overviews of the usage of custom optical projectors to study electronic crosstalk and astrometric distortions due to sensor effects during a period of focal plane testing using a subset of RTMs installed into the cryostat are presented in Section \ref{sec:electronic_crosstalk} and Section \ref{sec:astrometric_distortions} respectively. Finally, Section \ref{sec:conclusions} consists of a summary of the results of the electro-optical characterizations presented in this study as well as a discussion of future improvements and extensions that are planned for the upcoming testing of the full LSST Camera focal plane.

\section{FOCAL PLANE TESTING}\label{sec:focal_plane_testing}

\begin{figure}[htb]
	\begin{center}
		\begin{tabular}{c}
			\includegraphics[height=9cm]{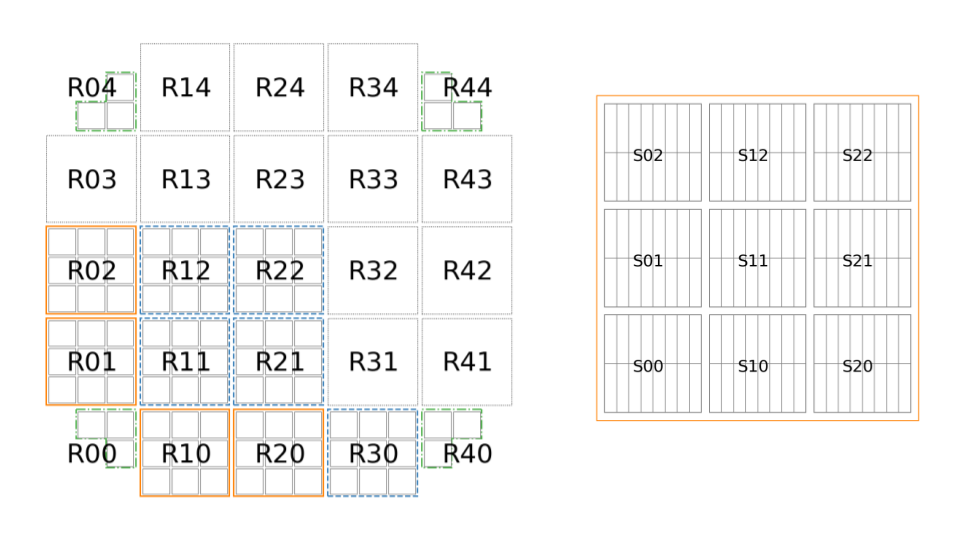}
		\end{tabular}
	\end{center}
	\caption 
	{ \label{fig:raft_layout}
		Left: Diagram of LSST Camera focal plane during the nine-raft testing period, showing ITL RTMs (orange solid outline), Teledyne e2v RTMs (blue dashed outline), corner RTMs (green dash-dotted outline), and uninstalled RTMs (black dotted outlines) labeled by raft slot names. Right: Diagram of a single RTM showing nine CCDs, labeled by sensor slot names, as well as the orientation of the 16 CCD segments (gray)} 
\end{figure}

Focal plane testing refers to the testing of the CCDs after integration of the RTMs into the LSST Camera cryostat. This ongoing state of testing began with a period of operation of the focal plane subsystems with only a subset of RTMs installed, referred to as the partial focal plane testing period.  During the data taking discussed here there were nine science-grade RTMs (4 consisting of ITL CCDs and 5 consisting of Teledyne e2v CCDs) and the four CRTMs installed.  The partial focal plane testing period was important for verification of the operation of the RTMs within the focal plane, the refrigeration systems, the cryostat vacuum performance, the data acquisition software, and networking capabilities. An additional goal of this period was to identify electro-optical characteristics of the installed CCDs that would warrant further study for optimization or mitigation. The installation of the RTMs and electro-optical testing during the partial focal plane testing period was performed using a specialized test stand, referred to as the Bench for Optical Testing (BOT), in conjunction with a set of custom-built optical projectors designed to illuminate the focal plane with various light illumination patterns; this equipment will also be used for future testing of the full focal plane.

\subsection{Bench for Optical Testing}\label{sec:BOT}

\begin{figure}[htb]
\begin{center}
\begin{tabular}{c} 
\includegraphics[height=10cm]{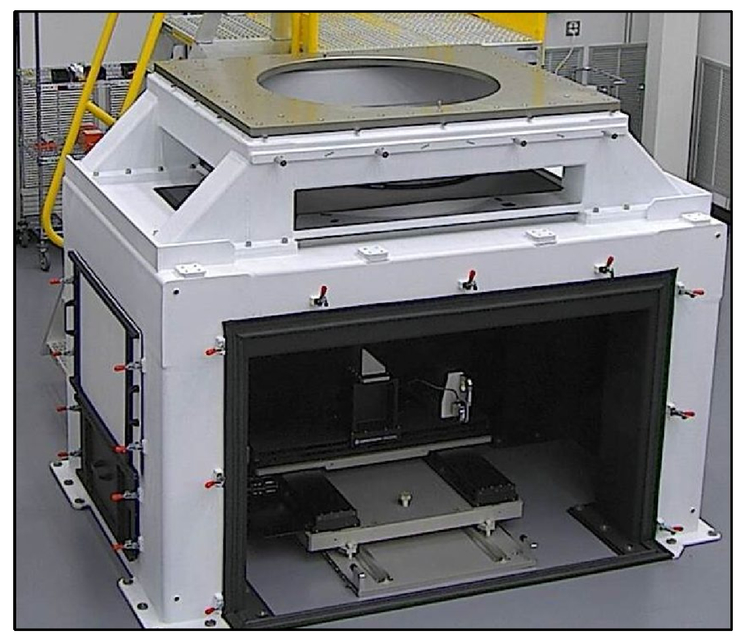}
\end{tabular}
\end{center}
\caption{ \label{fig:BOT_chassis} 
An image of the BOT Chassis with the dark box door panels removed. The LSST Camera cryostat is inserted into the circular opening of the mounting panel on the top.  The XY-motorplatform stages are visible below, in the dark box enclosure. Figure is taken from Newbry et al. \cite{Newbry2018}.}
\end{figure}

The BOT test stand, designed and built at SLAC, is important for verification testing of major Camera subsystems prior to full assembly of the LSST Camera. The BOT Chassis, pictured in Figure \ref{fig:BOT_chassis}, consists primarily of a large dark box enclosure that can be closed by a set of door panels to reduce background light to below 0.01 electrons per pixel per second.  The LSST Camera cryostat is inserted into the mounting panel, located at the top of the BOT, such that the focal plane is pointed downwards, to facilitate the installation of the RTMs. Located within the BOT dark box are two Aerotech PRO225 stages, referred to as the XY-motorplatform, along with a mounting interface that is used to position equipment beneath the focal plane. A detailed description of the BOT mechanical design and design requirements is provided by Newbry et al.\cite{Newbry2018}

\begin{figure}[htb]
\begin{center}
\begin{tabular}{c} 
\includegraphics[height=10cm]{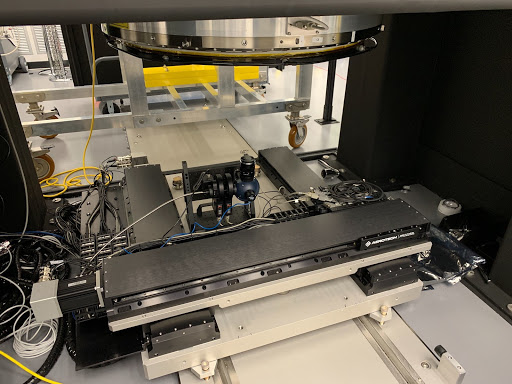}
\end{tabular}
\end{center}
\caption{ \label{fig:flat_illuminator} 
The flat illuminator projector positioned inside of the BOT, underneath the LSST Camera cryostat.}
\end{figure}

The primary projector for electro-optical testing of the focal plane is the \textit{flat illuminator projector}, shown positioned on the floor of the BOT dark box in Figure \ref{fig:flat_illuminator}. The flat illuminator projector uses a near-IR enhanced Zeiss 25mm f/2.8 wide-angle lens to re-image the 1.5" exit port of an integrating sphere in order to smoothly and fairly uniformly illuminate the entire focal plane. This optical design results in images that are sufficiently flat (less than 5\% signal variation across each CCD segment) to be used for measurements of the CCD non-linearity, full well, and gain from photon transfer curves. The light source for the flat illuminator projector is a Newport Fiber Optic Illuminator, containing a 150 W Quartz-Tungsten-Halogen lamp, that is delivered to the integrating sphere within the projector using a Newport single-branch glass fiber optic bundle connected to a Thorlabs 1" single-blade optical beam shutter. Two Thorlabs motorized filter wheels, connected in series between the shutter and the integrating sphere entrance port, are used to remotely control the light spectral bandpass and intensity. A set of Sloan Digital Sky Survey color filters (u, g, r, i, z, and Y) manufactured by Astrodon and a set of 6 narrow-band filters (10 nm full width at half maximum) centered at 480 nm, 650 nm, 750 nm, 870 nm, 950 nm, and 970 nm are installed in the first filter wheel. The second filter wheel holds a set of 6 masks with apertures of varying size that act as neutral density filters. A light baffle (not pictured in Figure \ref{fig:flat_illuminator}) is attached in front of the flat illuminator projector lens to limit the angular extent of the projected beam to eliminate reflected light from the inner side of the cryostat flange, which extends beyond the surface of the focal plane.

\begin{figure}[htb]
\begin{center}
\begin{tabular}{c} 
\includegraphics[height=10cm]{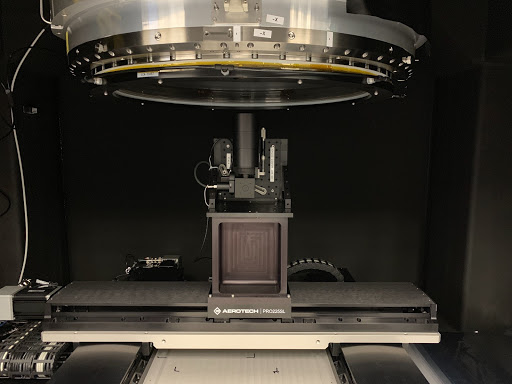}
\end{tabular}
\end{center}
\caption{ \label{fig:crosstalk_projector} 
The crosstalk projector mounted on the BOT XY-motorplatform underneath the LSST Camera cryostat during the partial focal plane testing period.}
\end{figure}

Electronic crosstalk measurements are made using a custom optical projector named the \textit{crosstalk projector} (Figure \ref{fig:crosstalk_projector}), that is designed to illuminate a single CCD with a pattern of large bright spots (80 pixels in radius) using collimated light. The decision to use an optical design that produced collimated light was made to reduce the impact of reflections of any projected bright spots that occur at the front and back side of the cryostat window, the surface of the CCD, and the back surfaces of the projector optics. Because collimated light will be perpendicular to these surfaces, the reflections of the bright spots will fall at the original location of the bright spot, rather than at different locations across the focal plane, which can interfere with the measurement of crosstalk. The light from the 1" exit port of an integrating sphere is passed through a pinhole, then through an aspheric lens, and finally through a metal mask to produce the desired pattern of bright spots. The mask used in the crosstalk projector was manufactured using precision wire electrical discharge machining and is painted matte black, to reduce its reflectivity. The light source is a QPhotonics 3 mW fiber-coupled 450 nm light emitting diode, connected by a single-mode fiber to a Thorlabs 1" single bladed optical beam shutter installed at the entrance port of the integrating sphere. When in use in the BOT, the crosstalk projector is mounted on a fixed stand on the XY-motorplatform, which is used to translate the projector in order to illuminate individual CCDs within the focal plane.

\begin{figure}[htb]
\begin{center}
\begin{tabular}{c} 
\includegraphics[height=10cm]{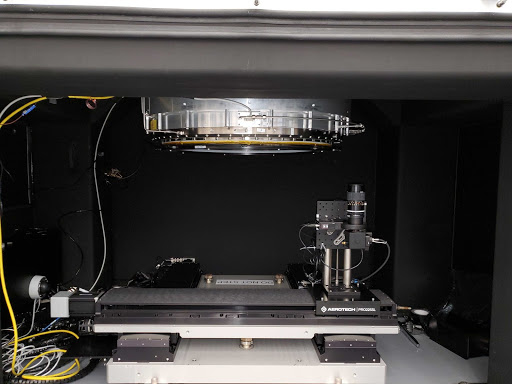}
\end{tabular}
\end{center}
\caption{ \label{fig:ch2_spot_projector} 
The spot projector mounted on the BOT XY-motorplatform underneath the camera cryostat during the partial focal plane testing period.}
\end{figure}

The ability to project realistically-sized sources onto the focal plane prior to on-sky operation is provided by a custom optical projector named the \textit{spot grid projector}.  The spot grid projector uses a Nikon 105mm f/2.8 Al-s Micro-Nikkor lens to re-image the 1" exit port of an integrating sphere that has been masked with the desired optical pattern using an HTA Photomask photo-lithographic mask. A six-position Thorlabs motorized filter wheel is used to remotely select from a set of masks with various patterns and position the desired mask in front of the integrating sphere exit port. During the partial focal plane testing period the optical patterns included in the spot grid projector filter wheel were: a single spot, a grid of 24x24 spots, a grid of 49x49 spots, a thin slit to simulate satellite streaks, and two absorptive neutral density filter wheels, which can be used to mimic a low-level sky background.  The spot grid projector shares the same 450 nm light emitting diode light source as the crosstalk projector and is also equipped with a Thorlabs 1" single-blade optical beam shutter. An additional feature of the spot projector is the ability to expose the focal plane to multiple illuminated mask patterns during a single exposure; this is accomplished by closing the shutter and rotating the filter wheel from one mask to another via remote command, before opening the shutter again, while continuing the CCD integration phase uninterrupted. When in use in the BOT, the spot grid projector is mounted on the XY-motorplatform using a manual adjustable z-stage. By adjusting the z-stage the spot grid projector can be positioned at the appropriate optical working distance to achieve good focus of the projected spots.

The optical projectors described above were designed to be modular to support rapid replacement and installation in the BOT during testing and are all controlled remotely using the same Camera Control System (CCS) software used to monitor and operate the other focal plane subsystems. In addition to the flat illuminator projector, the crosstalk projector, and the spot grid projector there exists an additional set of optical projectors that will not be discussed in depth but remain important for electro-optical testing. The Camera Calibration Optical Bench (CCOB) Wide-Beam projector, designed and built by collaborators at the Laboratory of Subatomic Physics \& Cosmology (LPSC) in Grenoble, France, is used to construct ``synthetic flat field images" from a number of co-added images that can be used to make quantum efficiency and pixel response non-uniformity measurements \cite{Newbry2018}.  Finally, a manually operated pinhole projector can be positioned underneath the focal plane to project images of printed pictures or physical objects for science outreach and publicity images, among other purposes \cite{Gnida2020}.

\subsection{Image Calibration}\label{sec:image_calibration}

The standard calibration for images taken during integration and testing of the LSST Camera CCDs includes an electronic offset correction, electronic bias correction, and dark current subtraction. The electronic offset correction is performed row by row; the mean signal of the serial overscan pixels in each is row (59 pixels total) is calculated, omitting the first 5 and last 2 overscan pixels, and then subtracted from the corresponding row of imaging pixels. The omission of the first 5 overscan pixels is done to reduce the contribution of serial deferred charge, while the omission of the last 2 overscan pixels is historically done as a result of large read noise in the final overscan pixels that was identified during earlier testing of individual LSST Camera CCDs. Electronic bias correction, using a superbias image constructed from a median stack of bias images, is necessary to remove additional spatial pixel non-uniformity across the image caused by the readout electronics and the rate of voltage clocking during readout. Dark current correction is done by subtracting an electronic offset and electronic bias-corrected superdark image, constructed from the median stack of multiple dark images, which is then scaled to match the exposure time of the image to be calibrated. For the analyses presented here, 10 bias images and 10 dark images were used when constructing the superbias and superdark images for calibration.  Gain values calculated from the photon transfer curve, which are in good agreement with gains calculated from \textsuperscript{55}Fe soft x-ray hits, are used for gain calibration. In some cases, these gains are further adjusted using an algorithm to scale each gain value in order to minimize the discontinuity of the pixel signals of a flat field image across segment boundaries, calculated using a least-squares minimization.

\section{ELECTRONIC CROSSTALK}\label{sec:electronic_crosstalk}

Electronic crosstalk refers to an effect in CCDs that have multiple channels that are read out simultaneously whereby a bright source that is present on one CCD segment will cause a ghost image to appear in the other CCD segments in the corresponding location, due to the electronic couplings between the output electronics. The electronic crosstalk between two CCD segments can be characterized by the crosstalk coefficient $c_{ij}$ defined as the ratio of the induced signal on a ``victim" segment $j$ compared to the true signal of the ``aggressor" segment $i$. Therefore to determine the true pixel signal of the victim segment $V_j$ one must subtract the pixel signal of the aggressor segment $A_i$ scaled by the crosstalk coefficient $c_{ij}$ from the measured pixel signals of the victim segment $V'_j$,

\begin{equation}\label{eq:crosstalk_correction}
V_j = V'_j - c_{ij} A_i\,.
\end{equation}

The heavily segmented nature of the LSST Camera focal plane (16 segments per each of the 189 science CCDs, not including the specialized CCDs used in the CRTMs) necessitates the characterization of the electronic crosstalk in three different regimes: within a single CCD, between CCDs in the same RTM, and between CCDs in different RTMs.  The electronic crosstalk is expected to be largest between segments located on the same CCD, referred to as intra-CCD crosstalk, where the primary electronic couplings are capacitive couplings between the on-chip readout electronics and off-chip trace-to-trace capacitive coupling within the cables connecting the CCDs to the external Readout Electronics Boards (REBs) that control the CCD operation and video channel output processing, including analog-to-digital conversion\cite{OConnor2015}.  The exact designs of the on-chip electronic readout chains of the CCDs manufactured by ITL and Teledyne e2v are not the same; therefore, there is no expectation that the behavior of the intra-CCD crosstalk will be identical between the two types of CCDs.

Crosstalk between segments on different CCDs located in the same RTM, referred to as intra-raft crosstalk could result from off-chip capacitive couplings that occur within or between REBs in a single RTM. A single REB is used to operate a row of three CCDs concurrently in an RTM; the same electronics are duplicated in three separate ``stripes" that are shielded from each other\cite{OConnor2016}. The presence of measurable intra-raft crosstalk would likely require hardware mitigation to improve the shielding and isolation of the individual CCD electronics within each REB board. 

The final regime of crosstalk is between segments on CCDs located on different RTMs, referred to as inter-raft crosstalk. Because this regime of crosstalk would need to be caused by electronic couplings after the analog-to-digital converters it is expected to be negligible. 

The goals of crosstalk measurements during the partial focal plane testing period were: verify the ability to measure crosstalk amplitude to the $10^{-6}$ level, determine the intra-CCD crosstalk behavior in a subset of ITL and Teledyne e2v CCDs, and to show that there is no intra-raft and inter-raft crosstalk at a significant enough level to warrant additional hardware mitigation.

\subsection{Crosstalk Measurement Method}\label{sec:crosstalk_methodology}

\begin{figure}[htb]
\begin{center}
\begin{tabular}{c} 
\includegraphics[height=11cm]{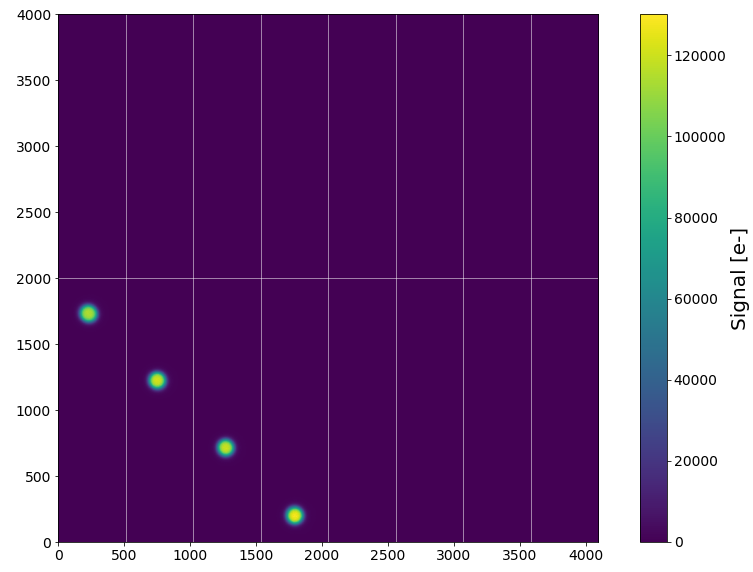}
\end{tabular}
\end{center}
\caption{ \label{fig:crosstalk_image} 
Example image taken using the crosstalk projector showing the illumination pattern of the four bright spots. The boundaries of the 16 segments are overlaid on this image (white) in order to demonstrate how each spot falls on a unique CCD segment and the corresponding victim regions are not co-located.}
\end{figure} 

Characterization of crosstalk for a single LSST Camera CCD, which has 16 segments, can be summarized by a 16x16 crosstalk matrix where each matrix element is the corresponding crosstalk coefficient $c_{ij}$. A single exposure taken using the crosstalk projector, shown in Figure \ref{fig:crosstalk_image}, results in an image with four bright spots illuminating four separate CCD segments.  The four spots are positioned such that the victim regions for each spot do not coincide with other victim regions or aggressor spots allowing for the simultaneous measurement of the crosstalk caused by four aggressor segments, corresponding to four rows in the crosstalk matrix. Using the BOT XY-motorplatform, the crosstalk projector is dithered to allow for the measurement of the crosstalk coefficients for all 16 CCD segments, using a minimum of four images. In order to make measurements of crosstalk values with magnitudes on the order of $10^{-6}$, the background noise must be reduced by taking multiple exposures per dither position (5 exposures per position during the partial focal plane test period) and generating a set of co-added images, calibrated using the procedures outlined in Section \ref{sec:image_calibration}. 

The spots produced by the crosstalk projector are not shaped as Gaussian functions or Airy disks; instead, the spots have a broad circularly symmetric center region approximately 80 pixels in radius, a full width at half maximum (FWHM) of approximately 150 pixels and no outer concentric rings.  The width of the spots is set by the diffraction that occurs at the photomask holes and increases as a function of the distance between the projector and the CCD surface. Diffraction at the photomask holes also results in diffraction spikes that contribute to background scattered light in the crosstalk images. During the partial focal plane testing period, the mean signal within a 80 pixel radius of the center of each spot was approximately 100,000 electrons; there is some evidence that the electronic crosstalk is dependent on the aggressor signal level, or non-linear behavior,\cite{Tyson2020} which motivates future study of this effect during the future focal plane testing.

The determination of each of the crosstalk coefficients is performed by fitting a crosstalk model to each victim region. Given a 200x200 pixel aggressor sub-image $A_i(m, n)$, where $m$ and $n$ are pixel indices, a corresponding 200x200 pixel victim sub-image $V_j(m, n)$ is modeled as

\begin{equation}\label{eq:crosstalk_victim}
V_j(m, n) = A_i(m, n) c_{ij} + mx_j + ny_j  + z_j\,,
\end{equation}
where $c_{ij}$ is the crosstalk coefficient and the parameters $x_j$, $y_j$, and $z_j$ define a sloped plane that represents the possible contribution of any additional background scattered light. The over-determined system of equations defined by Equation \ref{eq:crosstalk_victim} can be solved using an ordinary least-squares formalism.

\begin{figure}[htb]
\begin{center}
\begin{tabular}{c} 
\includegraphics[height=7cm]{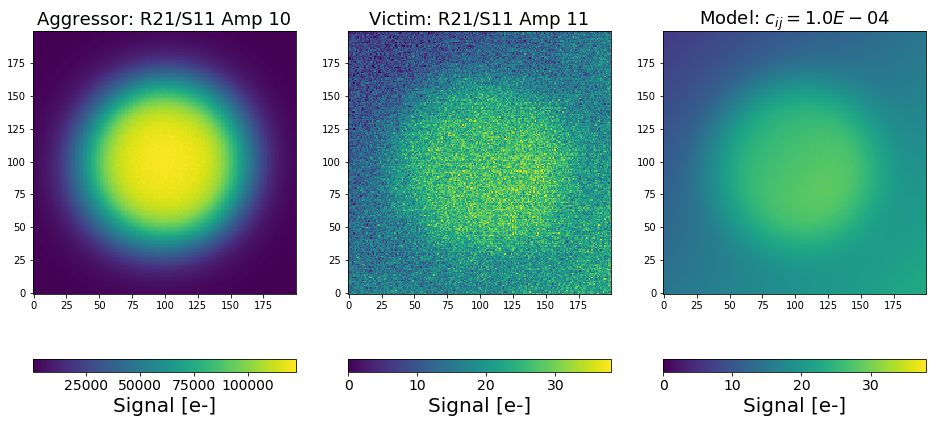}
\end{tabular}
\end{center}
\caption{ \label{fig:crosstalk_model_fit} 
Comparison of 200x200 pixel sub-images of an aggressor region (Left), a corresponding victim region (Center), and a best-fit crosstalk model determined using an ordinary least-squares method (Right).}
\end{figure}

Figure \ref{fig:crosstalk_model_fit} illustrates this method, comparing an aggressor sub-image to a victim sub-image and the best-fit model of the victim sub-image, defined by Equation \ref{eq:crosstalk_victim}. A low-level scattered light background from the crosstalk projector is visible in the victim sub-image; thus, it was necessary to include the additional $x_j$, $y_j$ and $z_j$ parameters to model this background and improve the accuracy of the crosstalk measurements beyond that of a simple ratio of signals between the victim and aggressor region. This methodology results in parameter estimate errors for the crosstalk coefficient that are approximately $10^{-6}$ or lower, for the majority of best-fit models.

\subsection{Crosstalk Characterization}\label{sec:ch4_crosstalk_characterization}

For visual clarity the crosstalk matrix results presented in this paper are plotted using a logarithmically binned color scale where each color bin represents a power of 10. Because electronic crosstalk can be both negative or positive, depending on the specific electronic couplings between segments, this color scale allows either positive or negative bins, symmetrically around 0. Crosstalk coefficient values whose amplitudes fall below a threshold of $|c_{ij}| < 5 \times 10^{-6}$, which corresponds to 1 electron of induced victim signal for a 200,000 electron aggressor signal, are plotted as white. Absent data and the crosstalk coefficients $c_{ii}$ that lie on the matrix diagonal are plotted as black. The naming convention in this paper is to label each RTM by its location in the focal plane (e.g. R00, R01, etc.), to label each CCD by its location within its RTM (e.g. S00, S01, etc.) and to label each segment by its output amplifier number (1 through 16), where output amplifiers 1 through 8 lie on one side of the CCD mid-line break and output amplifiers 9 through 16 lie on the opposite side.

\begin{figure}[htb]
\begin{center}
\begin{tabular}{c} 
\includegraphics[height=12cm]{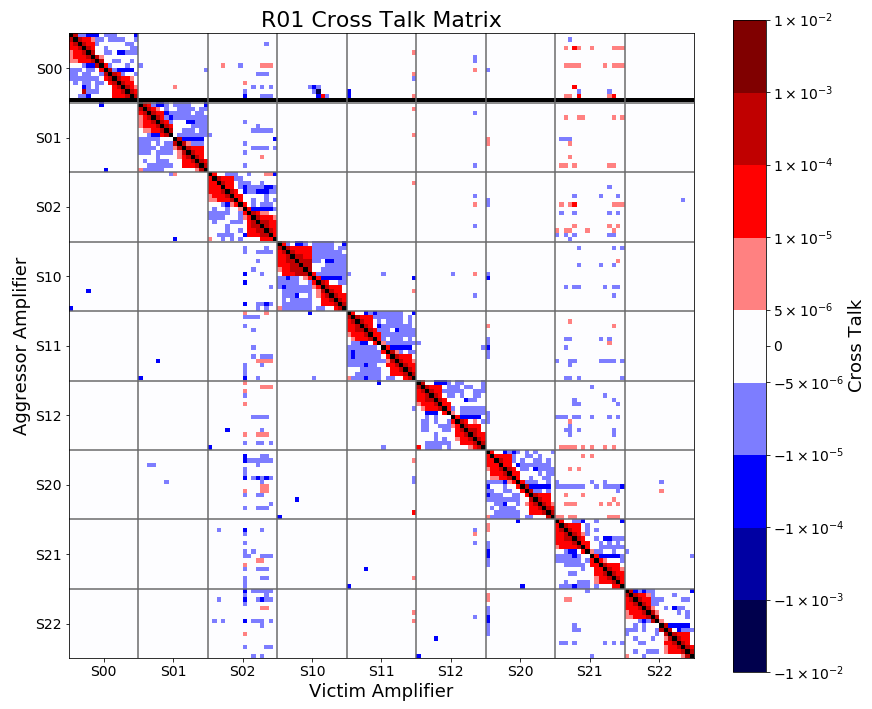}
\end{tabular}
\end{center}
\caption{ \label{fig:itl_raft_crosstalk_matrix} 
Crosstalk matrix for ITL RTM R01. Coefficient values have been binned logarithmically, in order to show the large range of magnitude of intra-CCD crosstalk. Crosstalk coefficient values whose amplitudes fall below a threshold of $|c_{ij}| < 5 \times 10^{-6}$ are plotted as white. Results colored black in the figure indicate the absence of a measurement.}
\end{figure}

The full 144x144 raft crosstalk matrix for ITL RTM R01 is shown in Figure \ref{fig:itl_raft_crosstalk_matrix}, where the 16x16 sub-matrices along the diagonal represent the intra-CCD crosstalk and the off-diagonal sub-matrices represent the intra-raft crosstalk between segments on different CCDs. The order of magnitude characterization of the intra-CCD crosstalk for each of the 9 CCDs in the RTM is largely consistent. The largest crosstalk is positive, on the order of $10^{-4}$, and is measured between segments that are immediately adjacent neighbors on the same side of the mid-line break while crosstalk measured between ITL CCD segments on opposite sides of the mid-line break is negative with magnitudes on the order of $10^{-5}$ and $10^{-6}$. Although the vast majority of the intra-raft crosstalk is below the threshold of $|c_{ij}| < 5 \times 10^{-6}$, there are patterns of larger crosstalk coefficients along columns of the matrix, corresponding to specific victim segments on CCDs such as R01/S02 and R01/S21. This pattern is not believed to be physical crosstalk, but instead results from elevated read noise in specific victim segments that increases the error on the parameter estimate calculations involving these segments. No crosstalk results were obtained for a single aggressor segment on CCD R01/S00 due to a surface defect on the cryostat window that blocked the projection of a bright spot on this segment during the crosstalk acquisition.

\begin{figure}[htb]
\begin{center}
\begin{tabular}{c} 
\includegraphics[height=12cm]{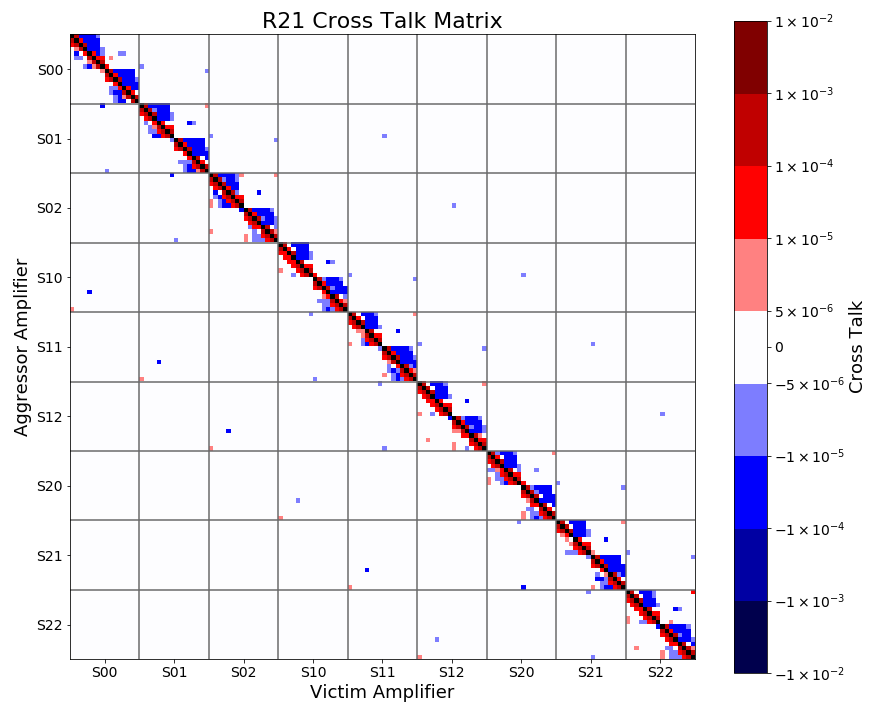}
\end{tabular}
\end{center}
\caption{ \label{fig:e2v_raft_crosstalk_matrix} 
Crosstalk matrix for Teledyne e2v raft R21. Coefficient values have been binned logarithmically, in order to show the large range of magnitude of intra-CCD crosstalk. Crosstalk coefficient values whose amplitudes fall below a threshold of $|c_{ij}| < 5 \times 10^{-6}$ are plotted as white. Results colored black in the figure indicate the absence of a measurement.}
\end{figure}

Figure \ref{fig:e2v_raft_crosstalk_matrix} shows the full 144x144 raft crosstalk matrix for Teledyne e2v RTM R21. The intra-CCD crosstalk behavior of Teledyne e2v CCDs is consistent between single CCDs but shows a greater abundance of negative crosstalk between segments on the same side of the mid-line break and much lower crosstalk between segments on opposite sides of the mid-line break, compared to ITL CCDs. The intra-CCD crosstalk matrices are also much less symmetric for Teledyne e2v CCDs than for ITL CCDs. The periodic, diagonal patterns seen in the intra-raft crosstalk measurements are caused by residual diffraction effects from the crosstalk projector optics, that repeat as the projector is dithered across the focal plane. Similar patterns can also be seen in the intra-raft crosstalk results for the ITL case, although they are largely subdominant to the previously discussed horizontal noise patterns along columns of the matrix. From these results, it was concluded that there was negligible intra-raft crosstalk for ITL or Teledyne e2v RTMs.

\begin{table}[htb]
\hspace{2em}
\caption{Science RTMs used for inter-raft crosstalk measurements for each of four possible combinations.} 
\label{tab:interraft_crosstalk_pairs}
\begin{center}       
\begin{tabular}{|c|c|c|} 
\hline
\rule[-1ex]{0pt}{3.5ex} Description & Aggressor RTM & Victim RTM \\
\hline\hline
ITL to ITL & R01 & R02 \\
\hline
ITL to Teledyne e2v & R01 & R11 \\
\hline
Teledyne e2v to Teledyne e2v & R21 & R22 \\
\hline
Teledyne e2v to ITL & R21 & R20 \\
\hline
\end{tabular}
\end{center}
\end{table}

\begin{figure} [htb]
\begin{center}
\begin{tabular}{c} 
\includegraphics[height=11cm]{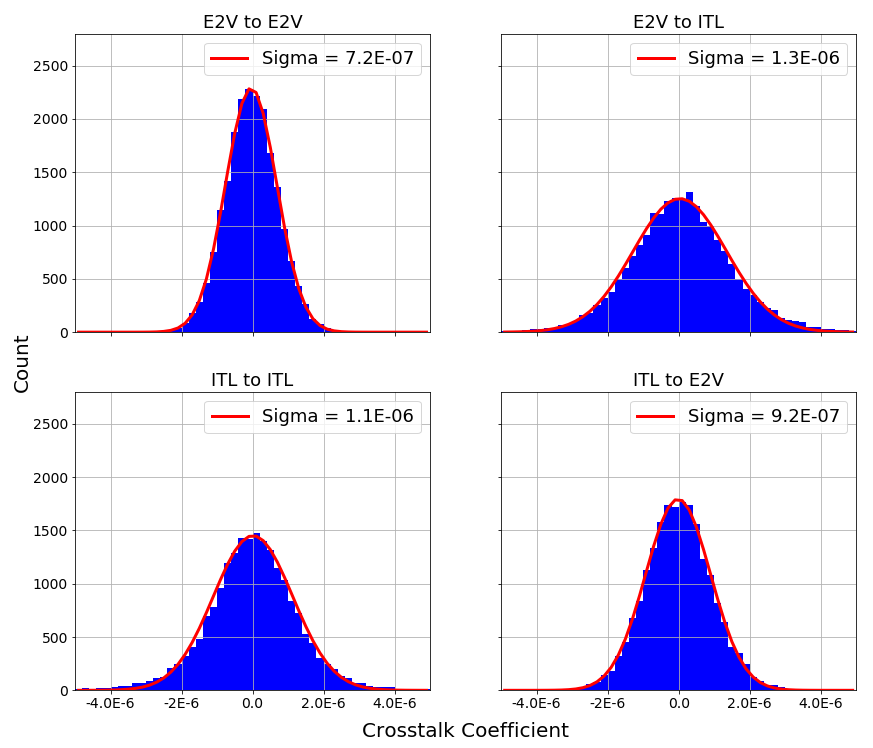}
\end{tabular}
\end{center}
\caption{ \label{fig:interraft_crosstalk_matrix} 
Histogram of crosstalk coefficient values and a chi-square fit Gaussian distributions for four cases of inter-raft crosstalk.}
\end{figure}

There are four cases of inter-raft crosstalk to consider, which were studied using a limited set of RTM pairings (Table \ref{tab:interraft_crosstalk_pairs}): crosstalk between two Teledyne e2v CCDs located on separate rafts, crosstalk between two ITL CCDs located on separate rafts, crosstalk between an aggressor ITL CCD and a victim Teledyne e2v CCD, and crosstalk between an aggressor Teledyne e2v CCD and a victim ITL CCD. In Figure \ref{fig:interraft_crosstalk_matrix} the distributions of crosstalk coefficients for each of the four cases are plotted and compared to a Gaussian distribution, calculated using a chi-square minimization. The standard deviation of the fitted Gaussian distributions is consistent with the standard error on the crosstalk coefficient parameter estimate, indicating that there is no measurable inter-raft crosstalk above measurement noise. The slightly larger widths of the distributions of inter-raft crosstalk involving RTMs consisting of ITL CCDs is attributed to the slightly elevated read noise characteristics of ITL CCDs compared to Teledyne e2v CCDs, a roughly 1 electron or 15 percent deviation.

\section{ASTROMETRIC DISTORTIONS}\label{sec:astrometric_distortions}

A number of sensor effects will modify the effective areas of the pixels in the rectilinear CCD array; the resulting spatial variations in pixel sizes can bias measurements of object astrometry, shape, and flux. These effects can be broadly categorized as static effects independent of pixel signal and dynamic effects that are dependent on pixel signal \cite{Astier2015}. Well-known examples of causes of static pixel size variations in deep-depletion CCDs are tree-ring patterns and CCD edge effects.  Tree-ring patterns are caused by circularly symmetric variations in dopant impurity concentrations that formed during the silicon boule growing process.\cite{Park2017, Park2020} CCD edge effects refer to changes to the transverse electric fields that define each pixel near the CCD perimeter, due to the influence  of the guard drain voltage.  The brighter-fatter effect is a well-studied example of a dynamic pixel size variation, where the repulsive effect of an increasing amount of collected charge within a pixel will decrease the effective pixel size resulting in bright sources that are systematically larger in width.\cite{Antilogus2014} In addition to these previously studied effects, during the integration and testing of the LSST Camera CCDs a number of possible newly recognized causes of pixel size variations have been observed in flat field images, the most prominent being tearing patterns in Teledyne e2v CCDs\cite{Juramy2019}. Although this tearing pattern, referred to as ``classical" tearing, has been largely eliminated by modifications to the operating voltages and clock timings of the Teledyne e2v CCDs, residual tearing patterns consisting of sharp gradients in pixel response near the edges of CCD segments, named ``divisadero" tearing, remain.

Examination of astrometric and photometric residuals from on-sky dithered images of star fields can be used to characterize the effect of pixel size variations resulting from many of these sensor effects\cite{Plazas2014}. During the focal plane testing period of the LSST Camera, the spot grid projector provided the capabilities to project a fake star field consisting of a grid of spots to perform similar characterization prior to the on-sky operation of the LSST Camera. This will allow for the possible development of mitigation schemes to be included in the data reduction pipelines, if necessary.

\subsection{Determination of Astrometric and Shape Residuals}\label{sec:ch5_residuals_calculations}

\begin{figure}[htb]
\begin{center}
\begin{tabular}{c} 
\includegraphics[height=11cm]{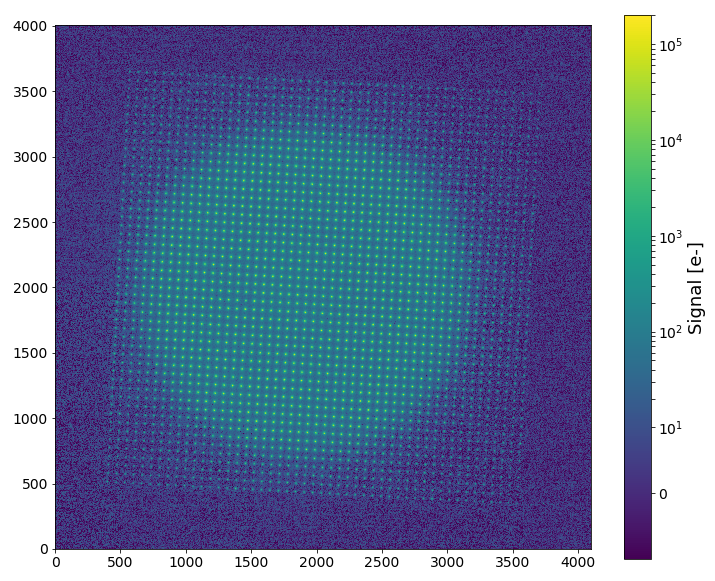}
\end{tabular}
\end{center}
\caption{ \label{fig:spot_grid_image} 
A calibrated CCD image of the 49x49 spot grid pattern projected by the spot projector. A logarithmic color scale that is linear between -10 and 10 electrons is used to show the range of magnitude of the different spot fluxes.}
\end{figure}

Figure \ref{fig:spot_grid_image} shows an image of a single CCD that has been illuminated using the 49x49 spot grid mask installed in the spot grid projector. Because spots near the edges of the grid are further from the optical axis and are greatly affected by the vignetting caused by the commercial lens, it is necessary to use a logarithmic color scale with a linear scaling between $\pm10$ electrons, in order to portray the range in magnitude of the spots in the entire grid. Spots located in the circular central region of the grid provide the largest signal-to-noise when measuring the astrometric and shape distortions caused by sensor effects.

Measurements of the changes to spot centroid position and shape distortions as a function of pixel location on a CCD are made by calculating the residuals between the measured properties of the observed spots and the properties of an ideal projected grid of spots that has been rotated and magnified. The residual vector $\Delta\vec{r}$ between the position of an ideal grid spot $\vec{s}$ and the position of the corresponding detected spot $\vec{d}$ is

\begin{equation}\label{eq:ch5_spot_residual}
\Delta\vec{r} = \vec{d} - \vec{s} = \Delta\vec{r}_\mathrm{optics} + \Delta\vec{r}_\mathrm{mask}+\Delta\vec{r}_\mathrm{sensor} + \vec{\epsilon}\,,
\end{equation}
where the first term represents optical distortions from the projector, the second term represents deviations of the mask pattern from the ideal, the third term represents the distortions caused by sensor effects dependent on the pixel X/Y location of the spot on the CCD, and the final term is the error term. The optical distortion component $\Delta\vec{r}_\mathrm{optics}$ and mask pattern deviations $\Delta\vec{r}_\mathrm{mask}$ for an individual spot will be determined only by its column and row position in the spot grid. Therefore, by calculating the mean centroid residual $\overline{\Delta\vec{r}}$ over a number of images of the spot located at different CCD pixel locations one can calculate this constant contribution to the spot's centroid residual:

\begin{equation}\label{eq:ch5_optic_distortions}
    \overline{\Delta\vec{r}} = \Delta\vec{r}_\mathrm{optics} + \Delta\vec{r}_\mathrm{mask}\,.
\end{equation}
Given a sufficiently large set of measurements spanning a large region of the CCD, the sensor effect distortions terms and the error term are assumed to average to zero. Then, the distortions caused by sensor effects at a particular pixel location on the CCD surface will be

\begin{equation}
    \Delta\vec{r}_\mathrm{sensor}\bigr|_{\vec{s} + \overline{\Delta\vec{r}}} = \Delta\vec{r} - \overline{\Delta\vec{r}}
\end{equation}
where the location of the spot is at the CCD position $\vec{s} + \overline{\Delta\vec{r}}$, the ideal position of the grid spot plus the mean centroid shift caused by the optics and mask variations. By dithering the projected 49x49 grid of spots across the CCD, measurements of the sensor effect distortions can be made at many different pixel positions using the same spot and at the same pixel position using different spots.  

The above formulation can also be applied to measurements of shape residuals of the spots, with only minor modifications. The shape measurements used in this study were intensity weighted second-moments $I_{ij}$, where the $i, j$ indices represent the x or y coordinate. The second-moment distortions caused by sensor effects is calculated from the residual

\begin{equation}\label{eq:ch5_other_distortions}
    \Delta I_{ij, \mathrm{sensor}}\bigr|_{\vec{s} + \overline{\Delta\vec{r}}} = I_{ij} - \overline{I}_{ij} 
\end{equation}
after subtracting the mean second-moment value, where the mean is over a set of measurements spanning a large region of the CCD.

\subsection{Astrometric Distortion Measurements}\label{sec:ch5_distortion_measurements}

The method presented in the previous section was used to measure centroid shifts and shape distortions caused by sensor effects on two CCDs installed in the LSST Camera focal plane: Teledyne e2v CCD R22/S11 and ITL CCD R02/S02. The error terms for the spot centroid and second-moment measurements were first estimated from the standard deviation of the distributions of these measurements taken for 1000 images of the spot grid positioned at the exact same XY-motorplatform position. The approximate value of these errors are also summarized in Table \ref{tab:measurement_error}. The centroid and second-moment error terms for spots near the center of the spot grid are lower than for spots near the edge of the spot grid. There also is a slight asymmetry between measurements in the $x$ and $y$ directions, corresponding to the serial and parallel directions in the CCD pixel array. 

\begin{table}[!htb]
\caption{Summary of the approximate error on centroid and shape measurements for spots on the edge of the grid and spots in the center of the grid.} 
\label{tab:measurement_error}
\begin{center}       
\begin{tabular}{|c|c|c|c|} 
\hline
\rule[-1ex]{0pt}{3.5ex} Description & Measurement & Edge Spot Error & Center Spot Error\\
\hline\hline
X position & $x$ & $2.8 \times 10^{-2}$ [pixels] & $6 \times 10^{-3}$ [pixels]\\
\hline
Y position & $y$ & $1.8 \times 10^{-2}$ [pixels] & $6 \times 10^{-3}$ [pixels]\\
\hline
XX second-moment & $I_{xx}$ & $7.2 \times 10^{-2}$ [$\mathrm{pixels}^2$]& $6.2 \times 10^{-2}$ [$\mathrm{pixels}^2$]\\
\hline
YY second-moment & $I_{yy}$ & $5.7 \times 10^{-2}$ [$\mathrm{pixels}^2$]& $3.3 \times 10^{-2}$ [$\mathrm{pixels}^2$]\\
\hline
\end{tabular}
\end{center}
\end{table}

For each CCD, 1600 randomly dithered spot grid exposures were taken, where the X/Y positions of the XY-motorplatform used to dither the spot projector were determined by drawing from a uniform distribution centered at the center of each CCD and spanning $\pm5$ mm. Every spot grid image acquisition was taken with an exposure time of 20 seconds to ensure that the spots located in the center of the grid achieved peak brightness of approximately 100 k$e^-$ and calibration of the images followed the procedure outlined in Section \ref{sec:image_calibration}. Source identification and measurements were performed using version v18\textunderscore0\textunderscore0 of the Rubin Science Pipeline (\url{https://pipelines.lsst.io}) to generate the source catalogs for each of the spot grid images.\cite{Bosch2019}  The Rubin Science Pipeline uses an implementation of the Sloan Digital Sky Survey (SDSS) adaptive moments shape algorithm as one method to calculate the spot centroid and second-moments \cite{Bernstein2002}, which was chosen as the method to be used for the analyses presented in the following sections.

\begin{figure}[htb]
\begin{center}
\begin{tabular}{c} 
\includegraphics[height=12cm]{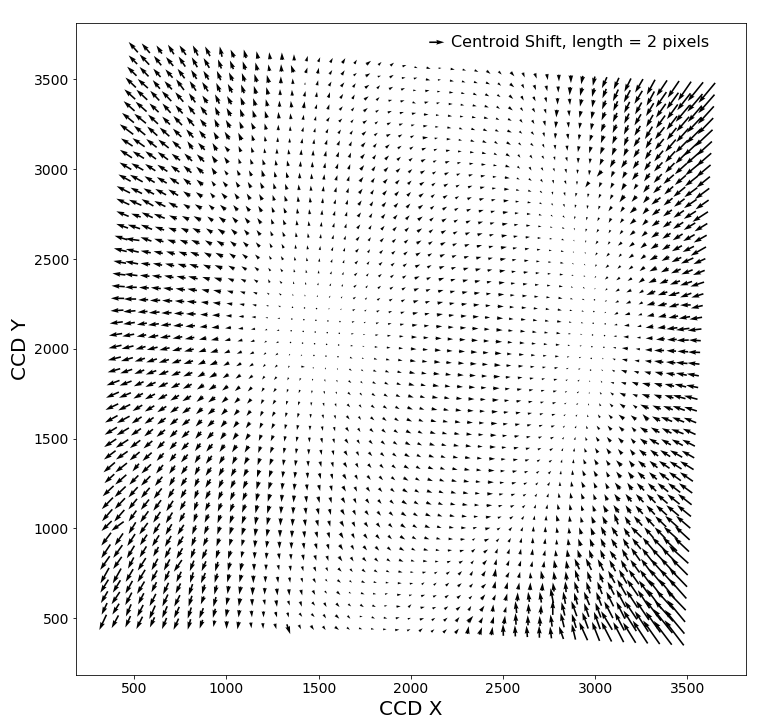}
\end{tabular}
\end{center}
\caption{ \label{fig:optical_centroid_shifts} 
Centroid shifts of each spot in the grid due to the projector optical distortions, presented as a vector field.}
\end{figure} 

The mean spot centroid residual $\overline{\Delta\vec{r}}$ for each of the spots in the grid representing centroid shifts caused by the projector optics and mask variations is shown in Figure \ref{fig:optical_centroid_shifts} as a vector field. Here, the grid is centered at the center of the CCD, each spot is plotted at the pixel location where the spot is expected to be imaged, and the displacement vector direction and magnitude correspond to the direction and magnitude of the mean centroid residual. Spots located near the edges of the grid are most affected by the optical distortions, with many experiencing centroid shifts of several pixels, from the ideal grid position. Because of the precision of the photolithographic process used to manufacture the mask, it is expected that any centroid shifts due to mask variations will be subdominant to the optical distortions. Although not visible at the scaling used in Figure \ref{fig:optical_centroid_shifts} there is some evidence for a systematic variation in the spacing between columns of the grid of spots on the photolithographic mask, likely resulting from the manufacturing patterning procedure.

The total quantity of data, due to the large number of total spots in the grid and number of exposures, allows for calculations of centroid and shape residuals caused by sensor effects at many points on the CCD surface. The results are presented as 1-to-1 ``pseudo-images"; at each CCD pixel coordinate, a Gaussian-weighted average (using a Gaussian with $\sigma_x= \sigma_y = 2$ pixels) of the residual measurements is calculated. Due to the size of the spot grid relative to that of the CCD and the extent of the random dither, the central region was well sampled using primarily the spots with the largest signal-to-noise ratio. The edge regions of the CCD, however, only have a small number of measurements associated with them taken with lower signal-to-noise and there are some regions of the pixel array where data was completely absent.

\begin{figure}[htb]
\begin{center}
\begin{tabular}{c} 
\includegraphics[height=8.2cm]{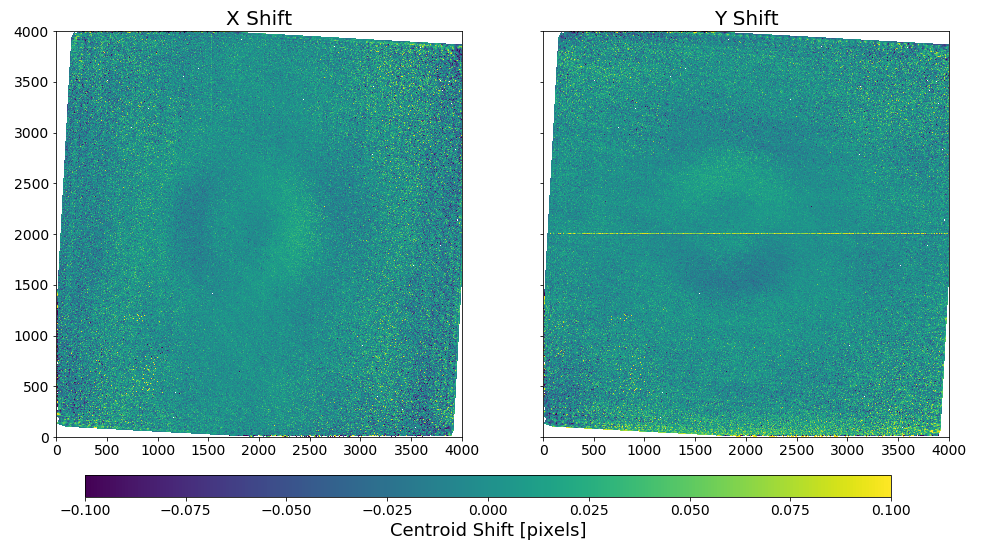}
\end{tabular}
\end{center}
\caption{ \label{fig:e2v_centroid_shifts} 
Centroid shifts of objects in the $x$ and $y$ directions due to sensor effects on Teledyne e2v CCD R22/S11.}
\end{figure} 
Figure \ref{fig:e2v_centroid_shifts} shows the centroid shifts in the X and Y directions caused by sensor effects on Teledyne e2v CCD R22/S11. The largest centroid shift measured for this CCD is the centroid shift in the y-direction of sources that fall on the mid-line break implant. Although difficult to see at this resolution there is also some indication of the effect of tree-ring patterns at a very low level. A notable feature seen in both the x-direction and y-direction centroid shift results is a large dipole feature located near the center of the CCD. The dipole feature appears visually similar to the central patterns of the optical centroid shifts (Figure \ref{fig:optical_centroid_shifts}), although the mean centroid residual of each spot, meant to remove the optical distortion component has been subtracted. This may indicate that it is necessary to calculate the per spot mean centroid residual from images of the grid that have been dithered over a larger portion of the CCD, in order to remove the effect of larger scale structure in the optical distortion mapping.  One possibility of a physical cause could be small pixel-area variations caused by strain on the silicon from the thermal, electrical and support connections of the sensor. Future testing of a larger number of CCDs of both types is necessary to determine if this dipole feature is inherent in Teledyne e2v CCDs or a result of a systematic effect in the residual calculations.

\begin{figure}[htb]
\begin{center}
\begin{tabular}{c} 
\includegraphics[height=8.2cm]{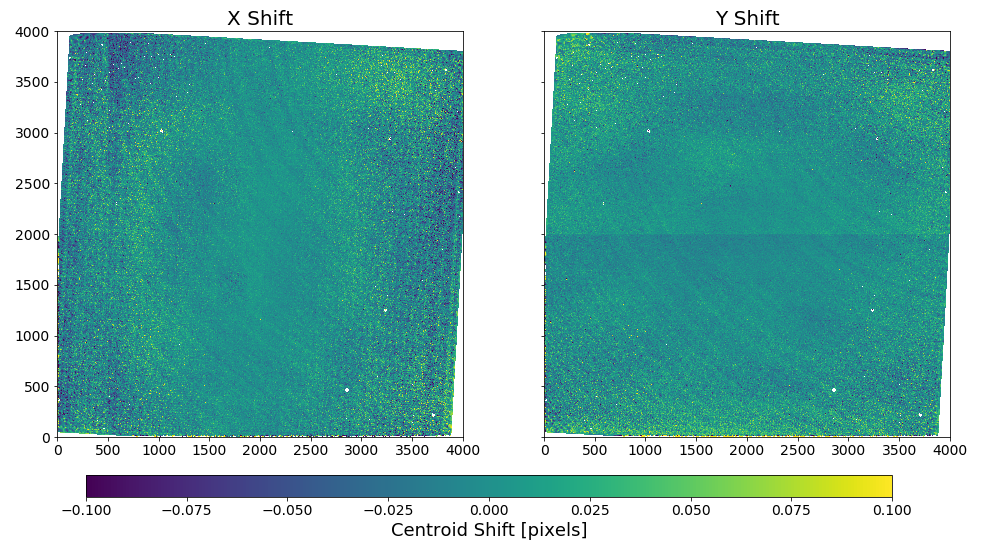}
\end{tabular}
\end{center}
\caption{ \label{fig:itl_centroid_shifts} 
Centroid shifts of objects in the $x$ and $y$ directions due to sensor effects on ITL CCD R02/S02.}
\end{figure} 
Figure \ref{fig:itl_centroid_shifts} shows the centroid shift in the X and Y directions caused by sensor effects on ITL CCD R02/S02. The tree-ring patterns are visible in both the x-direction and the y-direction centroid shifts, at a larger amplitude than in the Teledyne e2v CCD results, although still below $\pm0.1$ pixels. The effect of the mid-line break on centroid shifts in the y-direction is also much less pronounced. These results are consistent with pixel signal variations in the flat field images of ITL CCDs and Teledyne e2v CCDs and are caused by differences in the manufacturing and the operating conditions between the two types of CCDs. It is also notable that the dipole feature observed in the Teledyne e2v CCD results is not visible in the ITL CCD results, which provide some evidence that the dipole feature is caused by some property of the Teledyne e2v CCDs.

\begin{figure}[htb]
\begin{center}
\begin{tabular}{c} 
\includegraphics[height=8.2cm]{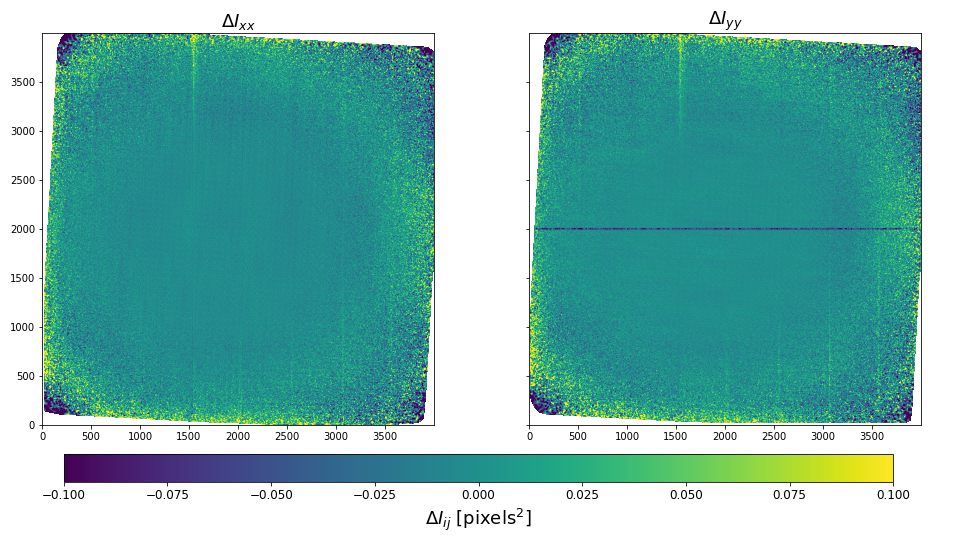}
\end{tabular}
\end{center}
\caption{ \label{fig:e2v_shape_distortions} 
Changes to the second-moments $I_{xx}$ and $I_{yy}$ of objects due to sensor effects on Teledyne e2v CCD R22/S11.}
\end{figure}

Accurate measurements of second-moment distortions caused by sensor effects are much more challenging due to the lower signal-to-noise; for this reason only the central region of the CCD has sufficient data and reduced measurement error to show trends above the noise. The results for Teledyne e2v CCD R22/S11 are shown in Figure \ref{fig:e2v_shape_distortions}. The effect of the mid-line break on the $I_{yy}$ second-moment is the strongest feature observed, causing changes of approximately $0.1$ $\mathrm{pixels}^2$. There are a number of horizontal streaks in both the $I_{xx}$ and $I_{yy}$ that appear to roughly follow the segment boundaries that are observed in shape distortion measurements, especially along the outer, higher noise regions of the shape distortion pseudo-images that may be signatures of shape distortions caused by divisadero tearing at the segment boundary. 

\begin{figure}[htb]
\begin{center}
\begin{tabular}{c} 
\includegraphics[height=8.2cm]{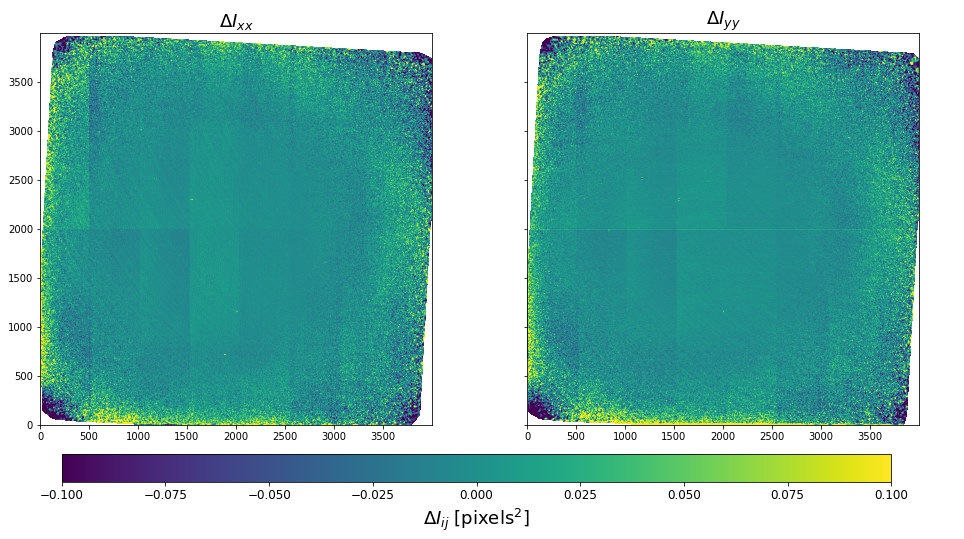}
\end{tabular}
\end{center}
\caption{ \label{fig:ch5_itl_shape_distortions} 
Changes to the second-moments $I_{xx}$ and $I_{yy}$ of objects due to sensor effects on ITL CCD R02/S02.}
\end{figure}

The second-moment distortion results for ITL CCD R02/S02, shown in Figure \ref{fig:ch5_itl_shape_distortions}, reveal prominent discontinuities between the different CCD segments. The exact cause of these discontinuities is not known at this time, though one possibility is that it is related to the strong serial deferred charge effects that have been measured in ITL CCDs, but are absent in Teledyne e2v CCDs.\cite{Snyder2019} Once again, the mid-line break is visible in the $I_{yy}$ second-moment results, but at a smaller amplitude than in the Teledyne e2v results.

\section{CONCLUSIONS}\label{sec:conclusions}

In this study, we have presented a number of studies related to electronic crosstalk and astrometric distortions that have been performed using a series of custom-built optical projectors during the partial focal plane testing period of the LSST Camera.  The results of these studies will be important for informing improvements and expansions to the planned testing and analyses to be performed during the upcoming full focal plane testing period.

The use of the custom-built crosstalk projector (Section \ref{sec:electronic_crosstalk}) during the partial focal plane test of the LSST Camera has allowed for characterization of the electronic crosstalk at the $10^{-6}$ level. The results presented in this study have shown that there is no detectable crosstalk between CCD segments located on different CCDs, either on the same RTM (intra-raft) or on different RTMs (inter-raft). The largest intra-CCD crosstalk is between immediately adjacent CCD segments located on the same side of the mid-line break and the amplitude of this crosstalk lies below the allowed maximum value of 0.002. Although there are several differences in the intra-CCD crosstalk behavior between ITL and Teledyne e2v CCDs, there is consistent behavior of the crosstalk within each type of CCD. During the upcoming full focal plane testing of the LSST Camera intra-CCD crosstalk measurements will be made for each of the 189 science CCDs and each of the CCDs within the CRTMs. Future crosstalk characterization will also include measurements of intra-CCD crosstalk made for a subset of CCDs at a larger range of aggressor signal values, by adjusting the exposure time used for each crosstalk projector image. This will allow for a greater understanding of any non-linear crosstalk behavior, that will be important for the development and testing of improved algorithms for crosstalk correction.

The use of the custom-built spot grid projector has demonstrated the ability to study centroid shifts and shape distortions of sources caused by sensor effects prior to on-sky operation of the LSST Camera. This will allow for identification and characterization of sensor effects that may impact the measurement of source shapes and fluxes needed for weak lensing science. The results of the preliminary measurements presented in this study have shown the ability to measure the effect of tree-ring patterns and the mid-line break on source centroids and shapes. A number of features such as the dipole in Teledyne e2v CCD centroid shift measurements, the discontinuities between segments in ITL CCD second-moment measurements, and the horizontal streaks close to segment boundaries in Teledyne e2v CCD second-moment measurements have also been revealed that will require follow-up study to determine if these are physical features of the sensors.

A number of improvements are planned for the upcoming full focal plane testing period: improving the dither pattern by drawing random dither positions from multiple Gaussian distributions positioned at different quadrants of the CCD (to more uniformly cover the CCD with the central spots of the grid), performing smaller dithers around specific regions of interest on a CCD that have shown large pixel response non-uniformities in flat field images, extending the residual analysis to include measurements of spot fluxes and the combined moment $I_{xx}+I_{yy}$, and increasing the number of CCDs of both manufacturers with a particular focus on CCDs that exhibit strong pixel response non-uniformities in flat field images. By replacing the 450 nm light emitting diode with light sources in other wavelengths, it will be possible to study the wavelength dependence of many of the sensor effects detailed here. The ability to project realistically sized sources is also planned to be leveraged to characterize brighter-fatter effect and signal-dependent deferred charge effects and to exercise the current correction algorithms in the data reduction pipelines.

\acknowledgments

This material is based upon work supported in part by the National Science Foundation through Cooperative Agreement 1258333 managed by the Association of Universities for Research in Astronomy (AURA), and the Department of Energy under Contract No. DE-AC02-76SF00515 with the SLAC National Accelerator Laboratory. Additional Rubin Observatory funding comes from private donations, grants to universities, and in-kind support from LSSTC Institutional Members.

We thank Eric Rosenberg for his contribution to the testing and development of the flat illuminator projector. We thank Laurel Doyle and Elisa Tabor for their contributions to the testing and development of the spot grid projector and centroid shift analysis algorithms. We thank And\'{e}s Plazas for reviewing this manuscript and providing valuable feedback.

\bibliography{main} 

\begin{thebibliography}{10}

\bibitem{Ivezic2008}
{Ivezi{\'c}}, {\v{Z}}., {Kahn}, S.~M., {Tyson}, J.~A., {Abel}, B., {Acosta},
  E., {Allsman}, R., {Alonso}, D., {AlSayyad}, Y., {Anderson}, S.~F., {Andrew},
  J., {Angel}, J. R.~P., {Angeli}, G.~Z., {Ansari}, R., {Antilogus}, P.,
  {Araujo}, C., {Armstrong}, R., {Arndt}, K.~T., {Astier}, P., {Aubourg},
  {\'E}., {Auza}, N., {Axelrod}, T.~S., {Bard}, D.~J., {Barr}, J.~D., {Barrau},
  A., {Bartlett}, J.~G., {Bauer}, A.~E., {Bauman}, B.~J., {Baumont}, S.,
  {Bechtol}, E., {Bechtol}, K., {Becker}, A.~C., {Becla}, J., {Beldica}, C.,
  {Bellavia}, S., {Bianco}, F.~B., {Biswas}, R., {Blanc}, G., {Blazek}, J.,
  {Bland ford}, R.~D., {Bloom}, J.~S., {Bogart}, J., {Bond}, T.~W., {Booth},
  M.~T., {Borgland}, A.~W., {Borne}, K., {Bosch}, J.~F., {Boutigny}, D.,
  {Brackett}, C.~A., {Bradshaw}, A., {Brand t}, W.~N., {Brown}, M.~E.,
  {Bullock}, J.~S., {Burchat}, P., {Burke}, D.~L., {Cagnoli}, G., {Calabrese},
  D., {Callahan}, S., {Callen}, A.~L., {Carlin}, J.~L., {Carlson}, E.~L.,
  {Chand rasekharan}, S., {Charles-Emerson}, G., {Chesley}, S., {Cheu}, E.~C.,
  {Chiang}, H.-F., {Chiang}, J., {Chirino}, C., {Chow}, D., {Ciardi}, D.~R.,
  {Claver}, C.~F., {Cohen-Tanugi}, J., {Cockrum}, J.~J., {Coles}, R.,
  {Connolly}, A.~J., {Cook}, K.~H., {Cooray}, A., {Covey}, K.~R., {Cribbs}, C.,
  {Cui}, W., {Cutri}, R., {Daly}, P.~N., {Daniel}, S.~F., {Daruich}, F.,
  {Daubard}, G., {Daues}, G., {Dawson}, W., {Delgado}, F., {Dellapenna}, A.,
  {de Peyster}, R., {de Val-Borro}, M., {Digel}, S.~W., {Doherty}, P.,
  {Dubois}, R., {Dubois-Felsmann}, G.~P., {Durech}, J., {Economou}, F.,
  {Eifler}, T., {Eracleous}, M., {Emmons}, B.~L., {Fausti Neto}, A.,
  {Ferguson}, H., {Figueroa}, E., {Fisher-Levine}, M., {Focke}, W., {Foss},
  M.~D., {Frank}, J., {Freemon}, M.~D., {Gangler}, E., {Gawiser}, E., {Geary},
  J.~C., {Gee}, P., {Geha}, M., {Gessner}, C. J.~B., {Gibson}, R.~R.,
  {Gilmore}, D.~K., {Glanzman}, T., {Glick}, W., {Goldina}, T., {Goldstein},
  D.~A., {Goodenow}, I., {Graham}, M.~L., {Gressler}, W.~J., {Gris}, P., {Guy},
  L.~P., {Guyonnet}, A., {Haller}, G., {Harris}, R., {Hascall}, P.~A., {Haupt},
  J., {Hernand ez}, F., {Herrmann}, S., {Hileman}, E., {Hoblitt}, J.,
  {Hodgson}, J.~A., {Hogan}, C., {Howard}, J.~D., {Huang}, D., {Huffer}, M.~E.,
  {Ingraham}, P., {Innes}, W.~R., {Jacoby}, S.~H., {Jain}, B., {Jammes}, F.,
  {Jee}, M.~J., {Jenness}, T., {Jernigan}, G., {Jevremovi{\'c}}, D., {Johns},
  K., {Johnson}, A.~S., {Johnson}, M. W.~G., {Jones}, R.~L., {Juramy-Gilles},
  C., {Juri{\'c}}, M., {Kalirai}, J.~S., {Kallivayalil}, N.~J., {Kalmbach}, B.,
  {Kantor}, J.~P., {Karst}, P., {Kasliwal}, M.~M., {Kelly}, H., {Kessler}, R.,
  {Kinnison}, V., {Kirkby}, D., {Knox}, L., {Kotov}, I.~V., {Krabbendam},
  V.~L., {Krughoff}, K.~S., {Kub{\'a}nek}, P., {Kuczewski}, J., {Kulkarni}, S.,
  {Ku}, J., {Kurita}, N.~R., {Lage}, C.~S., {Lambert}, R., {Lange}, T.,
  {Langton}, J.~B., {Le Guillou}, L., {Levine}, D., {Liang}, M., {Lim}, K.-T.,
  {Lintott}, C.~J., {Long}, K.~E., {Lopez}, M., {Lotz}, P.~J., {Lupton}, R.~H.,
  {Lust}, N.~B., {MacArthur}, L.~A., {Mahabal}, A., {Mand elbaum}, R.,
  {Markiewicz}, T.~W., {Marsh}, D.~S., {Marshall}, P.~J., {Marshall}, S.,
  {May}, M., {McKercher}, R., {McQueen}, M., {Meyers}, J., {Migliore}, M.,
  {Miller}, M., {Mills}, D.~J., {Miraval}, C., {Moeyens}, J., {Moolekamp},
  F.~E., {Monet}, D.~G., {Moniez}, M., {Monkewitz}, S., {Montgomery}, C.,
  {Morrison}, C.~B., {Mueller}, F., {Muller}, G.~P., {Mu{\~n}oz Arancibia}, F.,
  {Neill}, D.~R., {Newbry}, S.~P., {Nief}, J.-Y., {Nomerotski}, A., {Nordby},
  M., {O'Connor}, P., {Oliver}, J., {Olivier}, S.~S., {Olsen}, K., {O'Mullane},
  W., {Ortiz}, S., {Osier}, S., {Owen}, R.~E., {Pain}, R., {Palecek}, P.~E.,
  {Parejko}, J.~K., {Parsons}, J.~B., {Pease}, N.~M., {Peterson}, J.~M.,
  {Peterson}, J.~R., {Petravick}, D.~L., {Libby Petrick}, M.~E., {Petry},
  C.~E., {Pierfederici}, F., {Pietrowicz}, S., {Pike}, R., {Pinto}, P.~A.,
  {Plante}, R., {Plate}, S., {Plutchak}, J.~P., {Price}, P.~A., {Prouza}, M.,
  {Radeka}, V., {Rajagopal}, J., {Rasmussen}, A.~P., {Regnault}, N., {Reil},
  K.~A., {Reiss}, D.~J., {Reuter}, M.~A., {Ridgway}, S.~T., {Riot}, V.~J.,
  {Ritz}, S., {Robinson}, S., {Roby}, W., {Roodman}, A., {Rosing}, W.,
  {Roucelle}, C., {Rumore}, M.~R., {Russo}, S., {Saha}, A., {Sassolas}, B.,
  {Schalk}, T.~L., {Schellart}, P., {Schindler}, R.~H., {Schmidt}, S.,
  {Schneider}, D.~P., {Schneider}, M.~D., {Schoening}, W., {Schumacher}, G.,
  {Schwamb}, M.~E., {Sebag}, J., {Selvy}, B., {Sembroski}, G.~H., {Seppala},
  L.~G., {Serio}, A., {Serrano}, E., {Shaw}, R.~A., {Shipsey}, I., {Sick}, J.,
  {Silvestri}, N., {Slater}, C.~T., {Smith}, J.~A., {Smith}, R.~C., {Sobhani},
  S., {Soldahl}, C., {Storrie-Lombardi}, L., {Stover}, E., {Strauss}, M.~A.,
  {Street}, R.~A., {Stubbs}, C.~W., {Sullivan}, I.~S., {Sweeney}, D.,
  {Swinbank}, J.~D., {Szalay}, A., {Takacs}, P., {Tether}, S.~A., {Thaler},
  J.~J., {Thayer}, J.~G., {Thomas}, S., {Thornton}, A.~J., {Thukral}, V.,
  {Tice}, J., {Trilling}, D.~E., {Turri}, M., {Van Berg}, R., {Vanden Berk},
  D., {Vetter}, K., {Virieux}, F., {Vucina}, T., {Wahl}, W., {Walkowicz}, L.,
  {Walsh}, B., {Walter}, C.~W., {Wang}, D.~L., {Wang}, S.-Y., {Warner}, M.,
  {Wiecha}, O., {Willman}, B., {Winters}, S.~E., {Wittman}, D., {Wolff}, S.~C.,
  {Wood-Vasey}, W.~M., {Wu}, X., {Xin}, B., {Yoachim}, P., and {Zhan}, H.,
  ``{LSST: From Science Drivers to Reference Design and Anticipated Data
  Products},'' {\em The Astrophysical Journal}~{\bf 873},  111 (Mar. 2019).

\bibitem{Radeka2009}
Radeka, V., Frank, J., Geary, J.~C., Gilmore, D.~K., Kotov, I., O'Connor, P.,
  Takacs, P., and Tyson, J.~A., ``{LSST} sensor requirements and
  characterization of the prototype {LSST} {CCDs},'' {\em Journal of
  Instrumentation}~{\bf 4},  P03002--P03002 (Mar. 2009).

\bibitem{Doherty2014}
Doherty, P.~E., Antilogus, P., Astier, P., Chiang, J., Gilmore, D.~K.,
  Guyonnet, A., Huang, D., Kelly, H., Kotov, I., Kubanek, P., Nomerotski, A.,
  O’Connor, P., Rasmussen, A., Riot, V.~J., Stubbs, C.~W., Takacs, P., Tyson,
  J.~A., and Vetter, K., ``{Electro-optical testing of fully depleted CCD image
  sensors for the Large Synoptic Survey Telescope camera},'' in [{\em High
  Energy, Optical, and Infrared Detectors for Astronomy
  VI}{\nolinebreak\hspace{0.1em}]},  Holland, A.~D. and Beletic, J., eds.,
  {\bf 9154},  388 -- 404, International Society for Optics and Photonics, SPIE
  (2014).

\bibitem{Kotov2016}
Kotov, I.~V., Haupt, J., O'Connor, P., Smith, T., Takacs, P., Niel, H., and
  Chiang, J., ``Characterization and acceptance testing of fully depleted thick
  {CCDs} for the large synoptic survey telescope,'' {\em Proc. SPIE}~{\bf
  9915},  99150V--99150V--13 (2016).

\bibitem{Lopez2018}
Lopez, M., Marshall, S., Bond, T., Haupt, J., Johnson, T., Neal, H., O'Connor,
  P., Rasmussen, A., Roodman, A., Takacs, P., and Utsumi, Y., ``{Acceptance
  testing for LSST camera raft tower modules},'' in [{\em Ground-based and
  Airborne Instrumentation for Astronomy VII}{\nolinebreak\hspace{0.1em}]},
  Evans, C.~J., Simard, L., and Takami, H., eds.,  {\bf 10702},  745 -- 759,
  International Society for Optics and Photonics, SPIE (2018).

\bibitem{Newbry2018}
Newbry, S., Lange, T., Roodman, A., Reil, K., Bond, T., Rasmussen, A., Bowdish,
  B., Snyder, A., Rosenberg, E., and Lee, V., ``{LSST camera bench for optical
  testing: design, assembly, and preliminary testing},'' in [{\em Ground-based
  and Airborne Instrumentation for Astronomy VII}{\nolinebreak\hspace{0.1em}]},
   Evans, C.~J., Simard, L., and Takami, H., eds.,  {\bf 10702},  1553 -- 1571,
  International Society for Optics and Photonics, SPIE (2018).

\bibitem{Gnida2020}
Gnida, M., ``{Sensors of world’s largest digital camera snap first
  3,200-megapixel images at SLAC}.'' SLAC National Accelerator Laboratory, 8
  September 2020
  \url{https://www6.slac.stanford.edu/news/2020-09-08-sensors-world-largest-digital-camera-snap-first-3200-megapixel-images-slac.aspx}.
\newblock (Accessed: 12 December 2020).

\bibitem{OConnor2015}
O’Connor, P., ``{Crosstalk in multi-output CCDs for LSST},'' {\em Journal of
  Instrumentation}~{\bf 10},  C05010–C05010 (May 2015).

\bibitem{OConnor2016}
O'Connor, P., Antilogus, P., Doherty, P., Haupt, J., Herrmann, S., Huffer, M.,
  Juramy-Giles, C., Kuczewski, J., Russo, S., Stubbs, C., and Van~Berg, R.,
  ``Integrated system tests of the {LSST} raft tower modules,'' {\em Proc.
  SPIE}~{\bf 9915},  99150X--99150X--12 (2016).

\bibitem{Tyson2020}
Tyson, J.~A., Željko Ivezić, Bradshaw, A., Rawls, M.~L., Xin, B., Yoachim,
  P., Parejko, J., Greene, J., Sholl, M., Abbott, T. M.~C., and Polin, D.,
  ``{Mitigation of LEO Satellite Brightness and Trail Effects on the Rubin
  Observatory LSST},'' (2020).

\bibitem{Astier2015}
Astier, P., ``{An introduction to some imperfections of CCD sensors},'' {\em
  Journal of Instrumentation}~{\bf 10},  C05013–C05013 (May 2015).

\bibitem{Park2017}
Park, H., Nomerotski, A., and Tsybychev, D., ``Properties of tree rings in
  {LSST} sensors,'' {\em Journal of Instrumentation}~{\bf 12},  C05015--C05015
  (may 2017).

\bibitem{Park2020}
Park, H.~Y., Karpov, S., Nomerotski, A., and Tsybychev, D., ``{Tree rings in
  Large Synoptic Survey Telescope production sensors: its dependence on radius,
  wavelength, and back bias voltage},'' {\em Journal of Astronomical
  Telescopes, Instruments, and Systems}~{\bf 6}(1),  1 -- 11 (2020).

\bibitem{Antilogus2014}
{Antilogus}, P., {Astier}, P., {Doherty}, P., {Guyonnet}, A., and {Regnault},
  N., ``{The brighter-fatter effect and pixel correlations in CCD sensors},''
  {\em Journal of Instrumentation}~{\bf 9},  C03048 (Mar. 2014).

\bibitem{Juramy2019}
Juramy, C., Antilogus, P., Le~Guillou, L., and Sepulveda, E., ``Tearing and
  related field distortions in deep-depletion charge-coupled devices,'' {\em
  Journal of Astronomical Telescopes, Instruments, and Systems}~{\bf 5},  1
  (Sept. 2019).

\bibitem{Plazas2014}
Plazas, A.~A., Bernstein, G.~M., and Sheldon, E.~S., ``{On-Sky Measurements of
  the Transverse Electric Fields' Effects in the Dark Energy Camera CCDs},''
  {\em Publ. Astron. Soc. Pac.} ,  000–000 (July 2014).

\bibitem{Bosch2019}
{Bosch}, J., {AlSayyad}, Y., {Armstrong}, R., {Bellm}, E., {Chiang}, H.-F.,
  {Eggl}, S., {Findeisen}, K., {Fisher-Levine}, M., {Guy}, L.~P., {Guyonnet},
  A., {Ivezi{\'c}}, {\v{Z}}., {Jenness}, T., {Kov{\'a}cs}, G., {Krughoff},
  K.~S., {Lupton}, R.~H., {Lust}, N.~B., {MacArthur}, L.~A., {Meyers}, J.,
  {Moolekamp}, F., {Morrison}, C.~B., {Morton}, T.~D., {O'Mullane}, W.,
  {Parejko}, J.~K., {Plazas}, A.~A., {Price}, P.~A., {Rawls}, M.~L., {Reed},
  S.~L., {Schellart}, P., {Slater}, C.~T., {Sullivan}, I., {Swinbank}, J.~D.,
  {Taranu}, D., {Waters}, C.~Z., and {Wood-Vasey}, W.~M., ``{An Overview of the
  LSST Image Processing Pipelines},'' in [{\em Astronomical Data Analysis
  Software and Systems XXVII}{\nolinebreak\hspace{0.1em}]},  {Teuben}, P.~J.,
  {Pound}, M.~W., {Thomas}, B.~A., and {Warner}, E.~M., eds., {\em Astronomical
  Society of the Pacific Conference Series} {\bf 523},  521 (Oct. 2019).

\bibitem{Bernstein2002}
{Bernstein}, G.~M. and {Jarvis}, M., ``{Shapes and Shears, Stars and Smears:
  Optimal Measurements for Weak Lensing},'' {\em The Astronomical Journal}~{\bf
  123},  583--618 (Feb. 2002).

\bibitem{Snyder2019}
Snyder, A. and Roodman, A., ``{Investigation of deferred charge effects in
  Large Synoptic Survey Telescope ITL sensors},'' {\em Journal of Astronomical
  Telescopes, Instruments, and Systems}~{\bf 5}(4),  1 -- 7 (2019).

\end{thebibliography}
\bibliographystyle{spiebib} 

\end{document}